\documentclass [12 pt] {article}
\usepackage [margin= 2 cm, headheight=6pt,headsep=0.1in,heightrounded]{geometry}
\usepackage{verbatim}
\usepackage{indentfirst}
\usepackage[normalem]{ulem}
\usepackage[titletoc,title]{appendix}
\usepackage{soul}
\usepackage{bm}
\usepackage{float}
\usepackage{amssymb}
\usepackage{multicol}
\usepackage{mathtools}
\usepackage[dvipsnames]{xcolor}
\usepackage{titlesec}
\usepackage{wrapfig}
\usepackage{setspace}
\allowdisplaybreaks[2]
\usepackage{parskip}
\usepackage{ragged2e}
\usepackage{tikz}
\usepackage{tikz-3dplot}
\usepackage{amsmath}
\usepackage{afterpage}
\usepackage{parskip}
\usepackage{graphicx}
\usepackage{subfigure}
\usepackage{subcaption}
\usepackage{authblk}
\usepackage{abstract}
\usepackage{hyperref}
\hypersetup{colorlinks=true, linkcolor=blue, citecolor=blue, urlcolor=blue}
\counterwithin{equation}{section}
\usepackage{cleveref}
\usepackage{cite}

\title{Non-Gaussianity of Tensor Induced Density Perturbations}
\author[1,2,3]{Mariam Abdelaziz}
\author[4]{Pritha Bari} 
\author[5,6,7,8]{Sabino Matarrese}
\author[5,9,10]{Angelo Ricciardone\thanks{Email addresses:  \\ \href{mariamtarekmohamed.abdelaziz-ssm@unina.it}{mariamtarekmohamed.abdelaziz-ssm@unina.it}, \\ \href{prithabari@ibs.re.kr}{prithabari@ibs.re.kr}, \\ \href{sabino.matarrese@pd.infn.it}{sabino.matarrese@pd.infn.it}, \\ \href{angelo.ricciardone@unipi.it}{angelo.ricciardone@unipi.it}}}
\affil[1]{\textit{\small Scuola Superiore Meridionale, Largo San Marcellino 10, I-80138 Napoli, Italy}}
\affil[2]{\small \textit{INFN, Sezione di Napoli, Via Cinthia Edificio 6, I-80126 Napoli, Italy}}
\affil[3]{\small \textit{Department of Astronomy, Space Science and Meteorology, Faculty of Science, Cairo University, Giza 12613, Egypt}}
\affil[4]{\small \textit{Cosmology, Gravity, and Astroparticle Physics Group, Center for Theoretical Physics of the Universe, Institute for Basic Science (IBS), Daejeon, 34126, Korea}} 
\affil[5]{\small \textit{Dipartimento di Fisica e Astronomia “Galileo Galilei”, Universit\`a degli Studi di Padova, Via Marzolo 8, I-35131, Padova, Italy.}}
\affil[6]{\small \textit{INFN, Sezione di Padova, Via Marzolo 8, I-35131 Padova, Italy.}}
\affil[7]{\small \textit{INAF, Osservatorio Astronomico di Padova, Vicolo dell’Osservatorio 5, I-35122 Padova, Italy.}}
\affil[8]{\small \textit{Gran Sasso Science Institute,
Viale F. Crispi 7, I-67100 L’Aquila, Italy.}}
\affil[9]{\small \textit{Dipartimento di Fisica “Enrico Fermi”, Università di Pisa, Largo Bruno Pontecorvo 3, Pisa I-56127, Italy.}}
\affil[10]{\small \textit{INFN, Sezione di Pisa,  Largo Bruno Pontecorvo,    Pisa I-56127, Italy.}}

\date{}
\begin{document}
\maketitle
\begin{abstract}
 We investigate the non-Gaussianity of second-order matter density perturbations induced by primordial gravitational waves (GWs). These tensor-induced scalar modes arise from local fluctuations in the GWs energy density, which is quadratic in tensor perturbations. The resulting second-order density contrast follows a chi-squared distribution, naturally exhibiting significant non-Gaussianity. We compute the bispectrum of these tensor-induced scalar modes and analyze its dependence on various primordial GWs power spectra, including scale-invariant, blue-tilted, Gaussian-bump, and monochromatic sources. We find that the bispectrum shape is inherently sensitive to the underlying GWs spectrum by construction. In particular, Gaussian-bump and monochromatic sources produce a strong signal peaking in the equilateral configuration, similar to the effect of scalar-induced tensor modes. Our findings reveal a new way to probe primordial GWs via galaxy surveys and highlight a unique feature of tensor-induced density perturbations, otherwise mimicking linear ones on sub-horizon scales.
\end{abstract}
\newpage
\tableofcontents
\section{Introduction}\label{sec1:intro}
The inflationary paradigm \cite{Guth:1980zm} has long been considered a cornerstone of the standard cosmological model, offering solutions to its shortcomings and providing a mechanism to generate the initial conditions of the observable universe. Scalar perturbations, originating from quantum vacuum fluctuations in the inflaton field, form the seeds for structure formation. Meanwhile, tensor perturbations, or primordial Gravitational Waves (PGWs), arise from fluctuations in the metric tensor and leave their imprint primarily on the polarization pattern of the Cosmic Microwave Background (CMB) \cite{BICEP2:2018kqh, Kamionkowski:1996zd, Seljak:1996gy} and are traditionally considered to have a minimal imprint on the large-scale structure (LSS).

The Stochastic Gravitational-Wave Background (SGWB) predicted by inflation is considered a ``smoking gun'' of the inflationary era \cite{Starobinsky:1980te, Lyth:1998xn}. Unlike other cosmological messengers, GWs interact so weakly with matter that they travel across cosmic distances largely unscathed, preserving pristine information about their origins \cite{Guzzetti:2016mkm, Caprini:2018mtu, Boyle:2007zx, Inomata:2020lmk,Bartolo:2019oiq,Bartolo:2019yeu,LISACosmologyWorkingGroup:2022kbp}.
Detecting this background remains one of the most important goals in modern cosmology, with efforts focusing on indirect evidence in the B-mode polarization of the CMB, as well as direct detection through interferometers \cite{Maggiore:1999vm, Kamionkowski:2015yta, Guzzetti:2016mkm, Watanabe:2006qe, sakamoto2022probing, CMB-S4:2020lpa, Campeti:2020xwn, Flauger:2020qyi, LiteBIRD:2022cnt,Caporali:2025mum,Branchesi:2023mws,LISACosmologyWorkingGroup:2024hsc,Abac:2025saz}.

Pulsar Timing Array (PTA) collaborations-NANOGrav, EPTA/InPTA, PPTA, and CPTA, have reported evidence of signals consistent with an isotropic SGWB \cite{NANOGrav:2023hvm, EPTA:2015qep, EPTA:2023xxk, Xu:2023wog, Reardon:2023zen}. These results rely on the detection of quadrupolar correlations in pulsar timing residuals, aligning with the predicted Hellings \& Downs curve \cite{Hellings:1983fr}. Although these results are encouraging, more research is necessary to ascertain if the reported signal is predominantly cosmological or astrophysical in origin. The leading astrophysical interpretation attributes the signal to an incoherent superposition of GWs from a cosmic population of inspiraling supermassive black hole binaries (SMBHBs) \cite{Sesana:2008xk, Burke-Spolaor:2018bvk, NANOGrav:2020bcs}, whose collective spectral properties closely match key features of the PTA data. 
On the other hand, several cosmological models propose early-universe origins for the signal. For example, GWs produced during first-order phase transitions in the early universe are expected to generate a SGWB with distinctive spectral features \cite{Caprini:2018mtu} or scalar-induced gravitational waves that have been extensively studied through both analytical approaches and numerical simulations \cite{Matarrese:1997ay, Kohri:2018awv, Domenech:2021ztg, Inomata:2019zqy, Baumann:2007zm, Espinosa:2018eve,Perna:2024ehx,Kugarajh:2025rbt,LISACosmologyWorkingGroup:2025vdz}. Separating all of these possibilities will require additional observations and analyses.  If confirmed, it would mark a breakthrough in gravitational wave astronomy following the historic detection of GWs from a binary black hole merger \cite{LIGOScientific:2016aoc}, which has already driven significant advances in GW observational techniques and theoretical modeling.

While the majority of the cosmological community has focused on the previously mentioned observational windows for detecting PGWs, this work explores a novel approach that provides an alternative pathway to indirectly detect PGWs through their imprint on the LSS. Several theoretical frameworks have suggested that PGWs leave observable signatures in the LSS. One notable example is the ``tensor fossils'' effect, in which long-wavelength GWs induce local deviations from statistical isotropy by coupling to scalar modes \cite{Dai:2013ikl, Jeong:2012df, Masui:2010cz, Dimastrogiovanni:2014ina,Ricciardone:2017kre,Ricciardone:2016lym}. Other effects include the indirect influence of GWs on galaxy clustering and weak lensing shear \cite{Kaiser:1996wk, Jeong:2012nu, Schmidt:2012nw}.

The main focus of this work is the generation of second-order scalar perturbations by linear tensor modes, referred to as ``tensor-induced scalar modes''. These modes were first identified in \cite{Tomita:1967wkp, Matarrese:1997ay} and later investigated in detail in \cite{Bari:2021xvf, Bari:2022grh}, where their impact on the present-day matter power spectrum, particularly on large scales, was examined.  In \cite{Bari:2021xvf}, the authors found that a significant gravitational wave power spectrum can indeed leave an observable signature on the matter power spectrum.
The key feature of these second-order perturbations, induced by linear GWs, is that they are absent on super-horizon scales, unlike their linear counterparts. As a result, they do not contribute to CMB temperature anisotropies on large scales, though they may influence smaller scales.
On sub-horizon scales, the induced matter density contrast closely resembles the linear one and can effectively be viewed as a linear perturbation induced by the fluctuation in the gravitational radiation that vanishes outside the horizon. 
This distinctive signature opens up a promising pathway for the detection and constraining of GWs on scales that are currently poorly constrained. Additionally, it presents a valuable opportunity to enhance the matter power-spectrum estimations. By observing those gravitational wave-induced corrections to the density contrast, we can indirectly probe the presence of GWs. Even in the absence of a detection, these corrections could still place meaningful limits on their amplitude. Upcoming large-scale structure surveys, including Euclid \cite{Euclid:2019clj}, DESI \cite{DESI:2018ymu}, SPHEREx \cite{SPHEREx:2016vbo}, SKA \cite{SKA:2018ckk}, the Roman Space Telescope \cite{Rose:2021nzt}, and the Vera Rubin Observatory (LSST) \cite{LSSTDarkEnergyScience:2021ryz}, are exceptionally well-positioned to explore this potential. 

In this work, we focus on a post-inflationary mechanism for the generation of second-order scalar perturbations, wherein the source is only the fluctuation in the energy density of GWs on sub-horizon scales, thereby isolating the tensor-induced contribution. It is important to note, however, that in specific inflationary scenarios, such as multi-field models, some of which are considered here as viable models for generating a significant GWs background to source our effect \cite{Namba:2015gja,Dimastrogiovanni:2016fuu}, metric perturbations can back-react on the inflationary field dynamics, giving rise to mixed scalar-tensor primordial bispectra or trispectra. These additional degrees of freedom, while potentially significant, lie beyond the scope of the present analysis and are left for future work.
In a recent study, it was shown that scalar perturbations can be generated without even relying on a scalar field (the inflaton). In that framework, inflation is driven by a de Sitter space-time (dS), where gravitational waves arise from quantum vacuum fluctuations, and scalar fluctuations are generated through second-order effects from the tensor modes \cite{Bertacca:2024zfb}.

Tensor-induced scalar perturbations arise from local fluctuations in the energy density of GWs, which are quadratic in tensor perturbations. Assuming GWs are Gaussian, the resulting density perturbations follow a chi-squared distribution, inherently generating significant non-Gaussianity (NG). This intrinsic NG, also highlighted in \cite{Bari:2021xvf}, is the primary focus of our study and quantification in this work.

It is important to emphasize that the type of NG examined in this work differs from the commonly studied cases in the literature. Here, the non-Gaussian signal primarily arises in the density perturbations rather than in the gravitational potential, which is the typical focus of most studies \cite{Bartolo:2004if, Liguori:2010hx, Celoria:2018euj}. This scenario was first explored in \cite{Verde:2000vr}, where it was referred to as `\emph{Model A}'. Moreover, thanks to the Central Limit Theorem,  the NG in this context has a negligible impact on the NG in the CMB. 

The results we show in this paper reveal that inflationary models featuring a Gaussian-bump primordial GWs power spectrum, such as Axion-field inflation models \cite{Namba:2015gja, Dimastrogiovanni:2016fuu, Thorne:2017jft,Maleknejad:2016qjz,Maleknejad:2018nxz}, lead to a large non-Gaussian signal of the mentioned matter perturbations, with the bispectrum peaking in the equilateral configuration. For comparison, we also analyze cases with scale-invariant and blue-tilted power spectra.

This paper is organized as follows. In Section \ref{sec2:bg}, we review the derivation of the expression for the second-order density contrast. Section \ref{sec3:ps} revisits the calculation of the power spectrum of those tensor-induced scalar perturbations, while their bispectrum is derived in Section \ref{sec4:B}. The results are presented in Section \ref{sec5:result}, followed by a discussion and conclusions in Section \ref{sec6:Conc}. Additionally, the paper includes several appendices containing more detailed calculations.

\section{Background}\label{sec2:bg}
In this work, we build on the analysis presented in \cite{Bari:2021xvf}, where the authors derived the evolution equation of the tensor-induced second-order density perturbations and showed its contribution to the present-day matter power spectrum. We summarize the essential aspects of their approach and revisit the power spectrum calculation. 

Our analysis is set in a flat Friedmann-Lemaître-Robertson-Walker (FLRW) space-time, described by the metric $ds^2 = a^2(\tau)\left[-d\tau^2 + d\boldsymbol{x}^2\right]$, where $\tau$ denotes the conformal time and $a(\tau)$ represents the scale factor. Moreover, we choose to work in a collision-less $\Lambda$CDM model and fix the synchronous and comoving gauge, which can be rendered time-orthogonal due to the absence of pressure gradients in the matter sector \cite{Kodama:1984ziu}.
In this gauge, a perturbed flat FLRW metric becomes
\begin{align}
    ds^2 = a^2(\tau) \left( -d\tau^2 + \gamma_{ij}(\bm{x}, \tau) dx^i dx^j \right)\,,
\end{align}
where the spatial metric $\gamma_{ij}$ includes second-order scalar and linear tensor modes. We neglect linear scalar and vector modes since they are statistically independent of the tensor-sourced scalar modes and can, in principle, be set to zero manually. Then, the perturbed spatial metric looks like
\begin{align}
\gamma_{ij} = \left(1 - \phi^{(2)}\right)\delta_{ij} + \frac{1}{2}D_{ij}\chi^{||(2)} + \chi_{ij}^{(1)T}\,,
\end{align}
where $\chi_{ij}^{(1)T}$, from here on $\chi_{ij}$,  represents the linear transverse and traceless tensor modes, $\phi^{(2)}$ and $\chi^{||(2)}$ are the second-order tensor-sourced scalar modes, and $D_{ij} = \partial_i\partial_j - \frac{1}{3}\nabla^2 \delta_{ij}$.

Subtracting the isotropic Hubble flow from the covariant derivative of the fluid four-velocity, one can get the  peculiar velocity–gradient tensor $\theta^i_j $ which, thanks to the choice of the comoving gauge, is found to coincide with the extrinsic curvature of constant $\tau$ hypersurfaces $K^i_j$ \cite{Matarrese:1995sb, Matarrese:1997ay}
\begin{align}
\theta^i_j = a\, u^i_{;j} - \mathcal{H} \delta^i_j = -K^i_j = \frac{1}{2}\gamma^{i k}\gamma'_{k j}\,.
\end{align}
where $\mathcal{H} = a'/a$ is the conformal Hubble parameter and the prime denotes derivation w.r.t. conformal time. 
Now, we can write down the Raychaudhuri and continuity equations as \cite{Matarrese:1997ay,Ehlers:1961xww,Raychaudhuri:1953yv}
\begin{align}\label{raych}
     \theta' + \mathcal{H}\theta + \theta^i_j\theta^j_i + 4\pi G a^2 \bar{\rho}_m \delta = 0\,,\\\label{cont}
     \delta' + (1 + \delta)\theta = 0\,,
\end{align}
where $\theta$ is the peculiar volume expansion scalar, $\rho_m$ is the mean matter density, and $\delta = (\rho_m - \bar{\rho}_m)/\bar{\rho}_m$ is the matter density contrast.

Perturbing $\theta$ and $\delta$ up to the second order, then substituting in Eq. \eqref{raych} and Eq. \eqref{cont} we get
\begin{align}\label{raych2}
    \theta'^{(2)} + \mathcal{H}\theta^{(2)} + 2\theta^{(1)i}_j\theta^{(1)j}_i + 4\pi G a^2 \bar{\rho}_m \delta^{(2)} = 0\,,\\\label{cont2}
    \delta'^{(2)} + 2\delta^{(1)}\theta^{(1)} + \theta^{(2)} = 0\,.
\end{align}

Considering that the additive term $2\delta^{(1)}\theta^{(1)}$ can be neglected, as it does not depend on the tensor contribution, we can combine these equations to derive the evolution equation of $\delta^{(2)}$ as \cite{Bari:2021xvf}
\begin{equation}\label{master}
   {\delta^{(2)}}''+{\cal H}{\delta^{(2)}}'-4\pi G a^2 \bar \rho_{\rm m}\delta^{(2)}=\frac{1}{2} {\chi^{ij}}'{\chi_{ij}}'.
\end{equation}

On the left-hand side, Eq. \eqref{master} resembles the evolution equation for the linear density contrast. However, it features a quadratic source term composed of tensor modes on the right-hand side, representing fluctuations in the GWs radiation energy density, given by
$$\rho_{GW}= \frac{1}{32 \pi G a^2 } \langle \chi^{\prime ij}\chi^{\prime}_{ij} \rangle\,.$$  

Due to this quadratic source, the density modes acquire a significant intrinsic non-Gaussianity, which we quantify in this work.

The homogeneous and sourced solutions of (\ref{master}) were given in \cite{Bari:2021xvf} as
\begin{equation}
\delta_h^{(2)} = c_1(\mathbf{x}) D_+(\tau) + c_2(\mathbf{x}) D_-(\tau)\,,
\end{equation}
and
\begin{equation} \label{Ssol}
\delta_s^{(2)} = D_+(\tau) \int_0^\tau d\tilde{\tau} \frac{D_-(\tilde{\tau})}{W(\tilde{\tau})} \frac{1}{2} \chi^{\prime ij} \chi^{\prime}_{ij} 
- D_-(\tau) \int_0^\tau d\tilde{\tau} \frac{D_+(\tilde{\tau})}{W(\tilde{\tau})} \frac{1}{2} \chi^{\prime ij} \chi^{\prime}_{ij}\,,
\end{equation}
where \( D_+ \) and \( D_- \) are the linear growing and decaying homogeneous solutions, and 
\( W(\tau) \equiv D_-(\tau)D'_+(\tau) - D_+(\tau)D'_-(\tau) \) is the Wronskian. We observe from the sourced solution that $\delta^{(2)}$ evolves in time similarly to the standard linear density perturbations. However, since it is primarily sourced by linear GWs, this effect is significant only on sub-horizon scales, as GWs are frozen outside the horizon. This also ensures that our effect contributes to seed LSS formation without producing CMB anisotropies on large scales, unlike the linear ones.

Consequently, our primary focus will be on scalar perturbations that enter the horizon during the matter-dominated epoch. A detailed discussion on the treatment of time integration in (\ref{Ssol}), is provided in \cite{Bari:2021xvf, Bari:2022grh}.
The expression for the second-order density perturbations in a matter-dominated era then becomes
\begin{equation}
    \delta^{(2)}(\boldsymbol{x}, \tau)=\frac{\tau^{2}}{10} \int_{0}^{\tau} \frac{d \Tilde{\tau}}{\Tilde{\tau}} \chi_{i j}^{\prime} \chi^{\prime i j}-\frac{1}{10 \tau^{3}} \int_{0}^{\tau} d \Tilde{\tau}{\Tilde{\tau}}^{4} \chi_{i j}^{\prime} \chi^{\prime i j}\,.
\end{equation}

We move to Fourier space and write 
\begin{equation}
    \chi_{ij}\big(\bm{x},\eta\big)= \sum_{\sigma} \int \frac{d^3k}{\big(2\pi\big)^3} e^{i\bm{k}\cdot\bm{x}} \chi_\sigma \big(\bm{k}, \eta\big)\epsilon^\sigma_{ij}\big(\bm{\hat{k}}\big) \,,
\end{equation}
where $\epsilon^\sigma_{ij}(\boldsymbol{\hat{k}})$ are the polarization tensors [i.e. $\epsilon^\sigma_{ij}(\boldsymbol{\hat{k}}){\epsilon^{\sigma' ij}}(\boldsymbol{\hat{k}})= 2\delta_{\sigma \sigma'}$] for the two GWs polarizations $\sigma=+, \times$, and $\chi_\sigma (\boldsymbol{k}, \tau)$ is the GWs mode function which sources the scalar perturbations. For scales entering the Hubble radius during matter domination, we adopt the tensor transfer function \cite{Watanabe:2006qe} 
$$
\chi_\sigma (\boldsymbol{k}, \tau)= A_\sigma (\boldsymbol{k})\frac{3j_1(k\tau)}{k\tau}\,,
$$  
where $j_1$ is the spherical Bessel function of order one, and $A_\sigma \big(\boldsymbol{k}\big)$ is a stochastic zero-mean random field with the auto-correlation function
\begin{equation}\label{aa}
     \langle A_{\sigma_1}\big(\boldsymbol{k}_1\big) A_{\sigma_2}\big(\boldsymbol{k}_2\big)\rangle= \big(2\pi\big)^3 \delta^3\big(\boldsymbol{k}_1+\boldsymbol{k}_2\big) \delta_{\sigma_1 \sigma_2} \frac{2\pi^2}{k_1^3}\Delta^2_{\mathcal{\sigma}}\big(k_1\big)\,,
\end{equation}
where $\Delta^2_{\mathcal{\sigma}}(k)$ is the dimensionless power spectrum for each GWs polarization.

The Fourier transform of the product of the tensor fields becomes a convolution and the
density contrast reads
\begin{align} \label{bigdelta}
\delta^{(2)}(\boldsymbol{k}, \tau)= & \sum_{\sigma, \sigma^{\prime}} \int \frac{d^{3} p}{(2 \pi)^{3}} \epsilon_{i j}^{\sigma}(\hat{\boldsymbol{p}}) \epsilon_{i j}^{\sigma^{\prime}}(\widehat{\bm{k}-\bm{p}})\nonumber\\
& \times\left[\frac{\tau^{2}}{10} \int_{0}^{\tau} \frac{d \Tilde{\tau}}{\Tilde{\tau}} \chi_{\sigma}^{\prime}\left(\boldsymbol{p}, \Tilde{\tau}\right) \chi_{\sigma}^{\prime}\left(\boldsymbol{{k}-{p}}, \Tilde{\tau}\right)-\frac{1}{10 \tau^{3}} \int_{0}^{\tau} d \Tilde{\tau} \Tilde{\tau}^4 \chi_{\sigma}^{\prime}\left(\boldsymbol{p}, \Tilde{\tau}\right) \chi_{\sigma}^{\prime}\left(\boldsymbol{k}-\boldsymbol{p}, \Tilde{\tau} \right)\right]\,.
\end{align}

We confine ourselves to the standard practice of considering only the growing mode. The
final expression for the second-order density perturbation becomes
\begin{equation} \label{deltaexp}
\delta^{(2)}(\boldsymbol{k}, \tau)= \sum_{\sigma, \sigma^{\prime}} \int \frac{d^{3} p}{(2 \pi)^{3}} \chi_{\sigma}(\boldsymbol{p}) \chi_{\sigma^{\prime}}(\boldsymbol{k}-\boldsymbol{p}) \epsilon_{i j}^{\sigma}(\hat{\boldsymbol{p}}) \epsilon^{\sigma^{\prime} i j}(\widehat{\boldsymbol{k}-\boldsymbol{p}}) 
{\left[\frac{9 \tau^{2}}{10} \int_{0}^{\tau} \frac{d \Tilde{\tau}}{\Tilde{\tau}}  j_2 \left(p\Tilde{\tau}\right) j_2 \left(\left|\boldsymbol{k}-\boldsymbol{p}\right| \Tilde{\tau}\right)\right] }.
\end{equation}

In the next section, we recall the calculations of the two-point function of the induced \(\delta^{(2)}\) using Eq. \eqref{deltaexp}, focusing on the contribution of \(P_{\delta^{(2)}}\) to the present-day matter power spectrum, thereby confirming the result of \cite{Bari:2021xvf}. Subsequently, the same Eq. \eqref{deltaexp} will be employed to compute the three-point function (in Fourier space, the bispectrum) and quantify its intrinsic non-Gaussianity.

\section{Power-Spectrum}\label{sec3:ps}
As the results of \cite{Bari:2021xvf} are fundamental to our study, in this section we recall the calculation of the power spectrum of \(\delta^{(2)}\). According to the definition of dimensionless matter power-spectrum $\Delta^2_{\mathcal{\delta}}(k)$ we can write
\begin{equation}\label{dd}
    \langle \delta^{(2)}\left(\boldsymbol{k_1},\tau\right)\delta^{(2)}\big(\boldsymbol{k_2},\tau\big)\rangle= \big(2\pi\big)^3 \delta^3\big(\boldsymbol{k_1}+\boldsymbol{k_2}\big) \frac{2\pi^2}{k^3}\Delta^2_{\mathcal{\delta}^{(2)}}\big(k_1 \big)\,.
\end{equation}

The power spectrum for individual polarization mode is related to the GWs power spectrum by the relation,  $\Delta^2_{\sigma}(k)= (1/2)\Delta^2_T(k)$. Using Eq. \eqref{aa}, we get the following expression 
\begin{align}\label{PS}
\Delta_{\delta^{(2)}}^{2}\left(k\right) & =  \frac{81 \tau^{4} k^3}{800 \pi}    \int d^3 p \, f(k,p,\theta)\frac{\Delta_{T}^{2}\left(p\right) \Delta_{T}^{2}\left(\left|\boldsymbol{k}-\boldsymbol{p}\right|\right)}{p^3\left|\boldsymbol{k}-\boldsymbol{p}\right|^{3}} 
 \left[ \int_{0}^{\tau} \frac{d \tilde{\tau}}{\tilde{\tau}^3}  j_2 \left(p \tilde{\tau}\right) j_2 \left(\left|\boldsymbol{k}-\boldsymbol{p}\right| \tilde{\tau}\right)\right]^2 \,,
\end{align}
where $f(k,p,\theta)$ is the expression of the polarization tensors
\begin{align} \label{f}
 f\big(k,p,\theta\big)\equiv  \sum\limits_{\sigma,\sigma'}\epsilon^{\sigma'}_{ij}\big(\bm{\hat{p}}\big)\epsilon^{\sigma ij}\big(\bm{\widehat{k-p}}\big)\epsilon^{\sigma'}_{kl}\big(\bm{-\hat{p}}\big)\epsilon^{\sigma kl}\big(\bm{\widehat{-k+p}}\big)\,,
\end{align}
given and calculated explicitly in Appendix \ref{AppA}.
Here $\theta$ is the angle between $\bm{\hat{k}}$ and $\bm{\hat{p}}$.

To solve the integrals in Eq. \eqref{PS}, it is convenient to introduce the dimensionless variables $x=p/k$ and $y = |\boldsymbol{k - p}|/k$ \cite{Espinosa:2018eve} and a dimensionless time variable $\eta = k \Tilde{\tau} $ \cite{Bartolo:2007vp}  while the integration domain of $x$ and $y$ is given by \cite{Espinosa:2018eve} as $(x+y) \geq 1 \wedge (x+1) \geq y  \wedge (y+1)\geq x$.  Furthermore, the time integration in Eq. \eqref{PS} will be extended to infinity instead of integration up to the exact conformal time today to allow solving the time integration in terms of hypergeometric functions, an approximation that is valid since GWs decay after horizon entry, the effect from $\tau$ to $\infty$ will be minimal w.r.t the one from $0$ to $\tau$.

The dimensionless power spectrum is given in \cite{Bari:2021xvf} in terms of the dimensionless variables as
\begin{equation} \label{PSeq}
\Delta^2_{\delta^{(2)}}(k) = \left(\frac{9 \tau^2 k^2}{10}\right)^2 \int_0^\infty dx  \int_{|x-1|}^{x+1} dy  (xy)^{-2}  f(x, y)  \Delta^2_{T}(xk) \Delta^2_{T}(yk)
 \left( \int_0^\infty  \frac{d\eta }{\eta^3} j_2(x\eta) j_2(y\eta) \right)^2\,,
\end{equation}
where the function $f(x,y)$ reads
\begin{equation}
    f(x, y) = \frac{1}{16x^4 y^4} \left[ x^8 + \left( y^2 - 1 \right)^4 + 4x^6 \left( 7y^2 - 1 \right) + 4x^2 \left( y^2 - 1 \right)^2 \left( 7y^2 - 1 \right) + x^4 \left( 6 - 60y^2 + 70y^4 \right) \right]\,.
\end{equation}

Next is to evaluate the integral Eq. \eqref{PSeq} numerically for various input gravitational wave power spectrums and assess its contribution to the present-day matter power spectrum. We present the results of this part in Fig.~\ref{fig:Powerspectrum}.

\section{Bispectrum}\label{sec4:B}
In this section, we aim to quantify the non-Gaussianity of the second-order induced density perturbations by computing the bispectrum $B_{\delta^{(2)}}$, which is defined in terms of the three-point function as
\begin{equation} \label{Bisdef}
\left\langle\delta^{(2)} \left(\boldsymbol{k}_{1} , \tau \right) \delta^{(2)}\left(\boldsymbol{k}_{2},\tau\right) \delta^{(2)}\left(\boldsymbol{k}_{3},\tau\right)\right\rangle=(2 \pi)^{3} \delta^{(3)}\left(\boldsymbol{k}_{1}+\boldsymbol{k}_{2}+\boldsymbol{k}_{3}\right) B_{\delta^{(2)}}\left(k_{1}, k_{2}, k_{3}\right)\,.
\end{equation}

We use Eq. \eqref{deltaexp} to compute the three-point function such that in our case, we get 
\begin{align}
&\left\langle\delta^{(2)}\left(\boldsymbol{k}_{1}, \tau \right) \delta^{(2)}\left(\boldsymbol{k}_{2}, \tau \right) \delta^{(2)} \left(\boldsymbol{k}_{3}, \tau \right)\right\rangle = \left(\frac{9\tau^2}{10}\right)^3  \sum_{\sigma_{1} \sigma_{1}^{\prime}} \sum_{\sigma_{2} \sigma_{2}^{\prime}} \sum_{\sigma_{3} \sigma_{3}^{\prime}}\int \frac{d^{3} p_1}{(2 \pi)^{3}} \int \frac{d^{3} p_2}{(2 \pi)^{3}} \int \frac{d^{3} p_3}{(2 \pi)^{3}} \nonumber\\[1mm]
&\quad\times \epsilon_{ij}^{\sigma_{1}}\left(\boldsymbol{\hat{p}_{1}}\right) \epsilon^{\sigma_{1}^{\prime} ij}\left(\widehat{\boldsymbol{k_{1}- \boldsymbol{p}_{1}}}\right) \epsilon_{kl}^{\sigma_{2}}\left(\boldsymbol{\hat{p}_{2}}\right) \epsilon^{ \sigma_{2}^{\prime} kl}\left(\widehat{\boldsymbol{k_{2}- \boldsymbol{p}_{2}}}\right) \epsilon_{mn}^{\sigma_{3}}\left(\boldsymbol{\hat{p}_{3}}\right) \epsilon^{\sigma_{3}^{\prime} mn } \left(\widehat{\boldsymbol{k_{3}- \boldsymbol{p}_{3}}}\right) \nonumber\\[1mm]
&\quad\times\left(\int_{0}^{\tau} \frac{d \tilde{\tau}_{1}}{\tilde{\tau}^3_{1}} j_2 \left(p_{1}\tilde{\tau}_{1}\right) j_2 \left(\left|\boldsymbol{k}_{1}-\boldsymbol{p}_{1}\right| \tilde{\tau}_{1}\right)\right) \left(\int_{0}^{\tau} \frac{d \tilde{\tau}_{2}}{\tilde{\tau}^3_{2}} j_2 \left(p_{2} \tilde{\tau}_{2}\right) j_2 \left(\left|\boldsymbol{k}_{2}-\boldsymbol{p}_{2}\right| \tilde{\tau}_{2}\right)\right) \nonumber\\[1mm]
&\quad\times\left(\int_{0}^{\tau} \frac{d \tilde{\tau}_{3}}{\tilde{\tau}^3_{3}} j_2 \left(p_{3} \tilde{\tau}_{3}\right) j_2 \left(\left|\boldsymbol{k}_{3}-\boldsymbol{p}_{3}\right| \tilde{\tau}_{3}\right)\right) \nonumber\\[1mm]
&\quad\times\left\langle \chi_{\sigma_{1}}\left(\boldsymbol{p}_{1}\right) \chi_{\sigma_{1}^{\prime}}\left(\boldsymbol{k}_{1}-\boldsymbol{p}_{1}\right)
\chi_{\sigma_{2}}\left(\boldsymbol{p}_{2}\right) \chi_{\sigma_{2}^{\prime}}\left(\boldsymbol{k}_{2}-\boldsymbol{p}_{2}\right)
\chi_{\sigma_{3}}\left(\boldsymbol{p}_{3}\right) \chi_{\sigma_{3}^{\prime}}\left(\boldsymbol{k}_{3}-\boldsymbol{p}_{3}\right) \right\rangle \,. \label{threepnt}
\end{align}

The complete expression for the six-point function of the tensor perturbations in Eq. \eqref{threepnt} can be found in Appendix \ref{AppA}. We have eight non-vanishing Wick contractions, each contributing equally to the bispectrum, allowing us to compute a single contraction and multiply the result by eight. \cite{Espinosa:2018eve}.

\begin{figure}[t!]
    \centering
    \includegraphics[width= 9 cm]{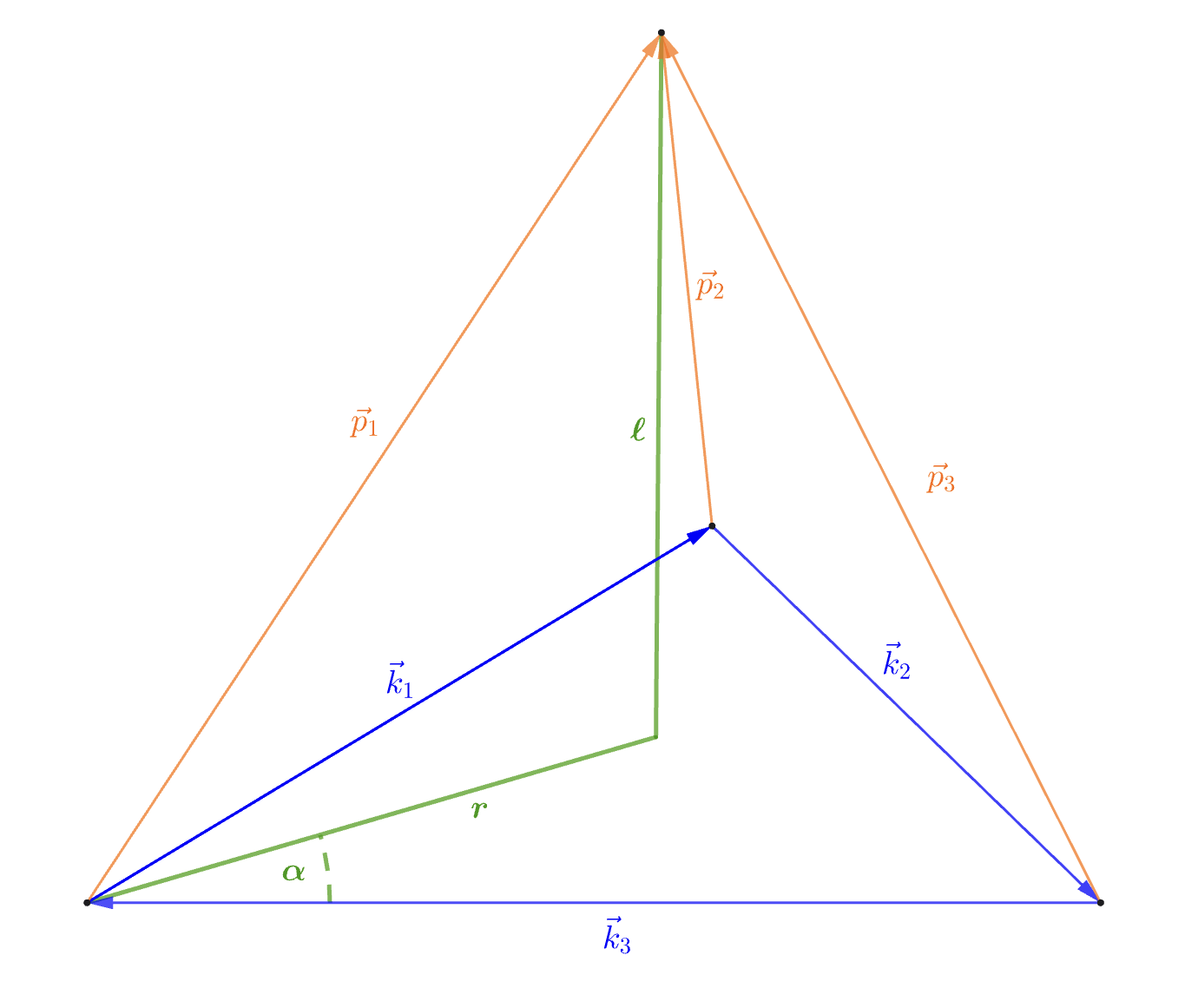}
    \caption{The geometrical configuration of the chosen contraction, a modified version of the figure from \cite{Espinosa:2018eve}.}
    \label{fig:1}
\end{figure}

The geometrical configuration of the six momenta, $\boldsymbol{k}_i$ and $\boldsymbol{p}_i$, shown in Fig.~\ref{fig:1}, is fixed by the Dirac deltas in Eq. \eqref{contr} \cite{Espinosa:2018eve}. We can analytically integrate over $ p_2$ and $p_3$ in Eq. \eqref{threepnt} using the two Dirac deltas in (\ref{contr}). The remaining integral over $p_1$ must be computed numerically, yielding the final expression for the bispectrum
\begin{align}
B_{\delta^{(2)}}(k_1,k_2,k_3) &= \; 0.48  \left(\frac{9 \pi \tau_o^2}{10}\right)^3 \int d^{3} p_{1} \,
\sum_{\sigma_{1} \sigma_{2} \sigma_3} \epsilon_{ij}^{\sigma_{1}}\left(\boldsymbol{\hat{p}_{1}}\right) \epsilon^{\sigma_{2}^* ij}\left(\boldsymbol{\hat{p}_2}\right) \epsilon_{kl}^{\sigma_{2}}\left(\boldsymbol{\hat{p}_{2}}\right) \epsilon^{ \sigma_{3}^{*} kl}\left(\boldsymbol{\hat{p}_3}\right) \epsilon_{mn}^{\sigma_{3}}\left(\boldsymbol{\hat{p}_{3}}\right) \epsilon^{\sigma_{1}^{*} mn}\left(\boldsymbol{\hat{p}_1}\right) \nonumber\\
& \times \left(\int_{0}^{\tau} \frac{d\tilde{\tau}}{\tilde{\tau}^3} 
j_2(p_1 \tilde{\tau})\, j_2(p_2 \tilde{\tau}) \right)
\left(\int_{0}^{\tau} \frac{d\tilde{\tau}}{\tilde{\tau}^3} j_2(p_2 \tilde{\tau})\, j_2(p_3 \tilde{\tau})\right)  \times \left(\int_{0}^{\tau} \frac{d\tilde{\tau}}{\tilde{\tau}^3} j_2(p_3 \tilde{\tau})\, j_2(p_1 \tilde{\tau})\right) 
\nonumber\\[2mm]
& \times \frac{\Delta^2_{T}(p_{1})}{p_{1}^{3}}\, \frac{\Delta^2_{T}(p_{2})}{p_{2}^{3}}\, \frac{\Delta^2_{T}(p_{3})}{p_{3}^{3}} \label{bispec}\,.
\end{align}

In Section \ref{sec2:bg}, we derived the expression for the induced density perturbation in a purely matter-dominated epoch. Incorporating the effects of dark energy leads to a different linear growth factor $D_+$ in Eq. \eqref{Ssol}, leading to the $0.48$ factor in Eq. \eqref{bispec}, which suppresses the matter growth. The fitting formula for the growth suppression factor for linear density perturbation is given in \cite{peacock_2010, Carroll:2000fy}.
Using $\Omega_m=0.32$ \cite{Planck:2018vyg}, we find our suppressed bispectrum to be $B_{\delta^{(2)}}(\Omega_m= 0.32, z =0)\simeq 0.48 B_{\delta^{(2)}}(\Omega_m= 1, z =0)$.

To solve the integral in (\ref{bispec}), we follow the procedure used in \cite{Espinosa:2018eve} and choose a cylindrical coordinate system for the integration variables such that
\begin{equation} \label{int}
\int \mathrm{d}^{3} p_1 \longrightarrow \int_{-\infty}^{+\infty} \mathrm{d} \ell \int_{0}^{+\infty} r \,\mathrm{d} r \int_{0}^{2 \pi} \mathrm{d} \alpha\,.
\end{equation}  

The new system of variables is shown in green in Fig.~\ref{fig:1}. We can also choose a frame such that 
\begin{equation} \label{ki}
\mathbf{k}_{1}=\left(k_{1x}, k_{1y}, 0\right), \quad \mathbf{k}_{2}=\left(k_{2x}, k_{2y}, 0\right), \quad \mathbf{k}_{3}=\left(-k_{3}, 0,0\right).
\end{equation}  

Consequently, the momenta $\boldsymbol{p_i}$ can be expressed as
\begin{equation} \label{pi}
\mathbf{p}_{1}=\left(r \cos \alpha, r \sin \alpha, \ell \right), \quad \mathbf{p}_{2}=\left(-k_{1x} + r \cos \alpha, -k_{1y} + r \sin \alpha, \ell \right), \quad \mathbf{p}_{3}=\left(-k_{3} + r \cos \alpha, r \sin \alpha, \ell \right).
\end{equation}

The explicit evaluation and final expression for the contractions of the polarization tensors in Eq. \eqref{bispec} are presented in Appendix \ref{AppB}. For the time integration, unlike in the case of the power spectrum, we integrate from \(0\) to the present time $\tau_{o}$, ensuring a more accurate result.  

Now, with Eqs. \eqref{int}, \eqref{ki}, \eqref{pi}, and \eqref{contraction}, we have all the necessary components to solve Eq. \eqref{bispec} numerically. The final step is to select a power spectrum model for the GWs and substitute it into Eq.~\eqref{bispec}. The results are presented in the next section.

\section{Results for different PGW sources}\label{sec5:result}
A diverse range of inflationary scenarios has been proposed, each predicting a distinct GWs power spectrum determined by its specific generation mechanism. While standard vacuum fluctuations of the gravitational field during inflation contribute to the spectrum, several well-motivated early Universe mechanisms can also produce a significant GWs background, particularly at small scales \cite{Barnaby:2010vf, Sorbo:2011rz}. In \cite{Bari:2021xvf}, the authors demonstrated that all models generating a strong monopole GWs signal could induce significant density perturbations, thereby affecting the matter power spectrum and producing a large NG.

Here, we consider various GWs power spectra, including some previously examined in \cite{Bari:2021xvf}, along with the nearly scale-invariant spectrum, a hallmark of the simplest inflationary models, as sources of density perturbations. 
The impact on the matter power spectrum is shown in Fig.~\ref{fig:Powerspectrum}. While for the results of the non-Gaussianity, we define a dimensionless normalized shape $S_{\delta^{(2)}}(k_1,k_2,k_3)$ \cite{Espinosa:2018eve}
\begin{equation}\label{Shape}
   S_{\delta^{(2)}}(k_1,k_2,k_3) = k_1^2 k_2^2 k_3^2 \frac{B_{\delta^{(2)}}(k_1,k_2,k_3)}{\sqrt{
   \Delta^2_{\delta^{(2)}}(k_1) \Delta^2_{\delta^{(2)}}(k_2) \Delta^2_{\delta^{(2)}}(k_3)}}\,.
\end{equation}

\begin{figure}  [t!]
    \centering
    \begin{subfigure}
             \centering
    \includegraphics[width=0.48\linewidth]{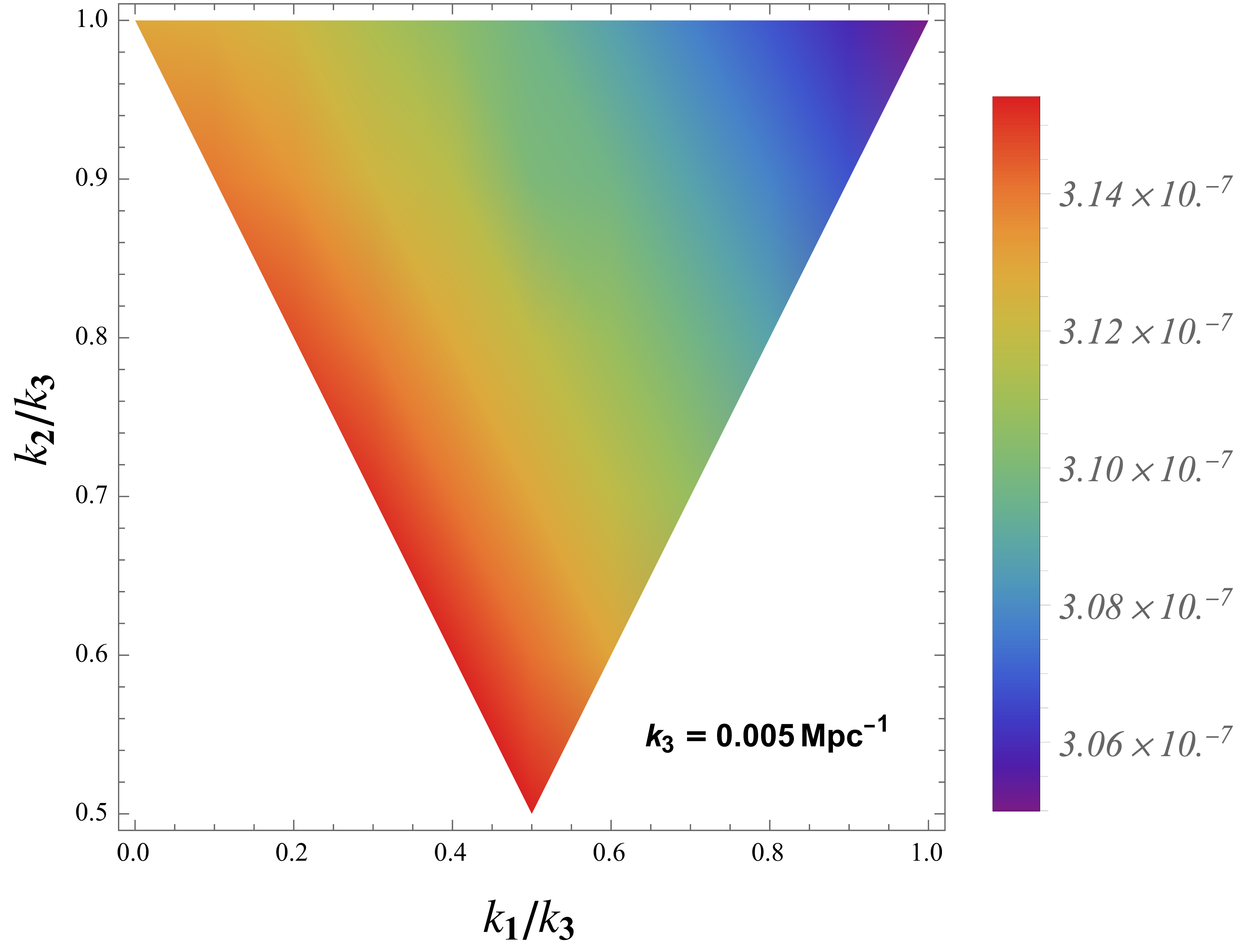}
    \end{subfigure}
    \hfill
    \begin{subfigure}
             \centering
    \includegraphics[width=0.48\linewidth]{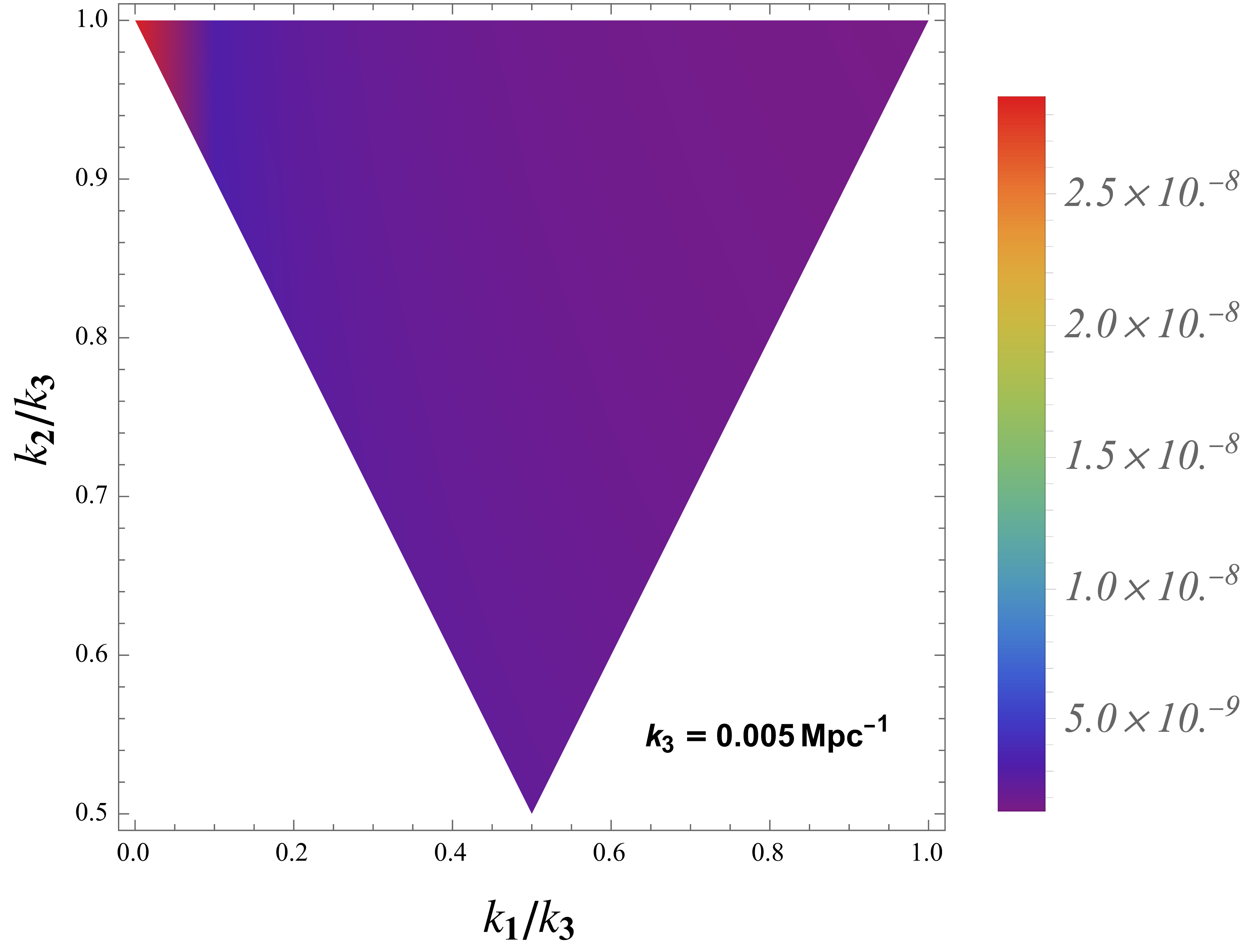}
    \end{subfigure}
    \caption{Normalized shapes for the bispectrum of the matter density contrast by scale invariant (left) and a power-Law (right) GWs sources fixing $k_3$ at $0.005 \text{ Mpc}^{-1} $.}
    \label{fig:combinedone}
\end{figure}

\begin{figure} [t!]
        \centering
\includegraphics[width=0.48\linewidth]{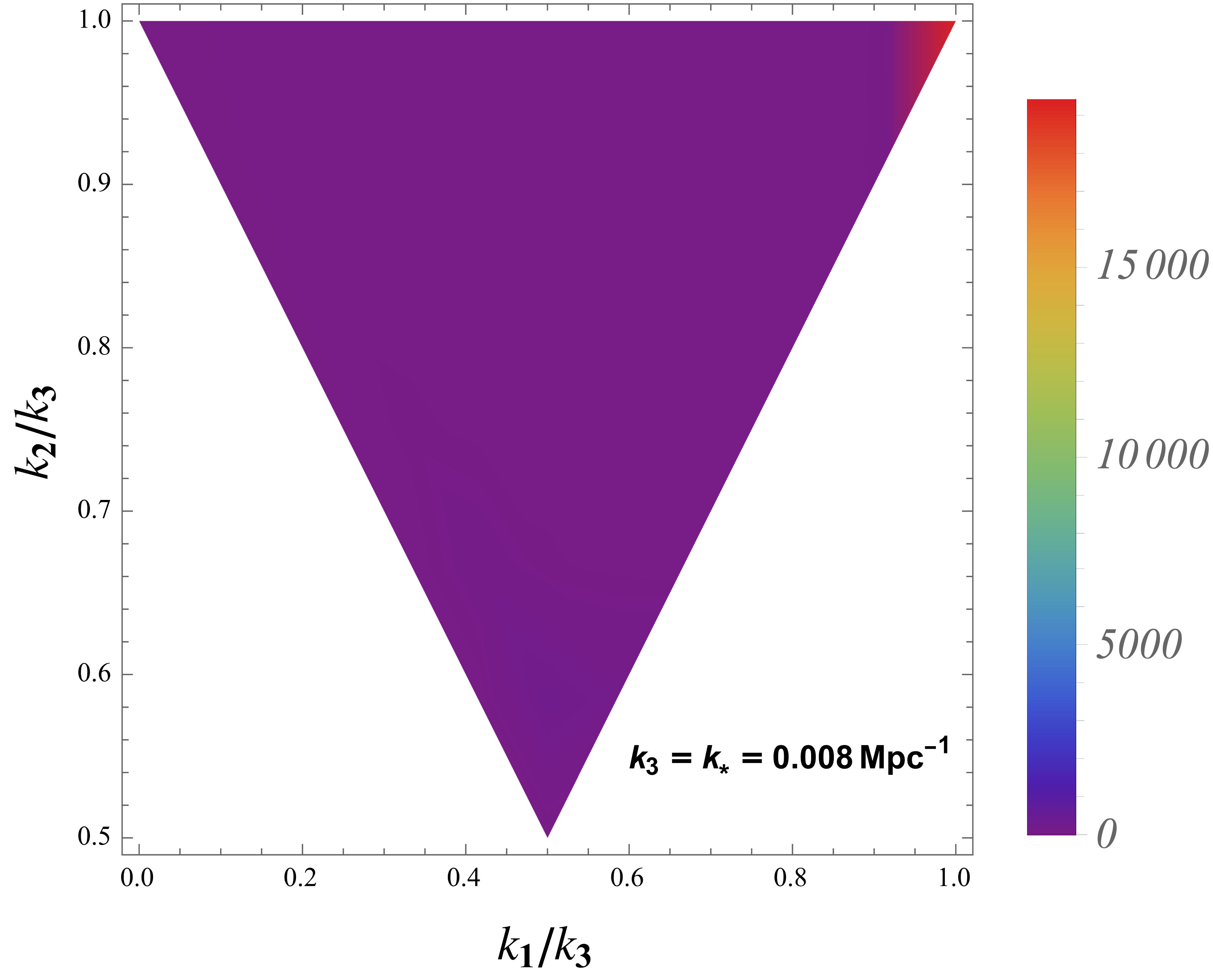}
    \hfill
    \caption{Normalized shape for the bispectrum of the matter density contrast by a monochromatic GWs source with $k_3$ fixed at $k_3 = k_* = 0.008 \text{ Mpc}^{-1}$.}
    \label{fig:Mono}
\end{figure}

We present the numerical results for the normalized bispectrum, keeping \( k_3 \) fixed and ordering the momenta as \( k_1 \leq k_2 \leq k_3 \). We plot $ S_{\delta^{(2)}}(k_1,k_2,k_3)$ for each of our selected GWs source models. The results are presented in \Cref{fig:combinedone,fig:Gauss,fig:Mono}.
\subsection{Scale invariant and power-law GW sources}
Power-law models of inflation predict a simple parametrization of the primordial tensor power spectrum, which can be expressed as  
\begin{equation}
\Delta^2_T(k) = A_{T} \left(\frac{k}{k_*}\right)^{n_T},
\end{equation}
where \( A_T \) is the amplitude of the tensor spectrum at a pivot scale \( k_* \), and \( n_T \) is the tensor spectral index. 

Standard single-field, slow-roll inflationary models predict a nearly scale-invariant power spectrum on super-horizon scales, characterized by  $n_T = -2\epsilon $, where  $\epsilon$ is the standard slow-roll parameter. 
The latest Planck data impose strong constraints on the tensor-to-scalar ratio on CMB scales, with $r_{0.01} < 0.066$ (at 95\%  Confidence Level (C.L.), PLANCK TT, TE, EE + lowE + lensing + BK15 + LIGO \& Virgo 2016), which restricts the range of the tensor spectral index to ( $-0.76 < n_T < 0.52 $) \cite{Planck:2018jri}. The blue tilt of GWs is particularly relevant for observational prospects at smaller scales, such as those probed by high-frequency GWs detectors, including pulsar timing arrays (e.g., NANOGrav) \cite{NANOGrav:2023gor, NANOGrav:2020bcs} and upcoming experiments like LISA \cite{Bartolo:2016ami}, making it a powerful candidate to include as a source in our analysis. In our case, we focus on a blue-tilted tensor power spectrum, selecting $n_T = 0.32$, which remains within the range permitted by current and forthcoming GWs interferometers \cite{Bartolo:2016ami, Planck:2018jri}. We also set the amplitude of the GWs power spectrum to $A_T=r A_s=0.06 \times 2.1 \times 10^{-9}=1.26\times 10^{-10}$. Our pivot scale is set at $k_*=k_{CMB}=0.01\, \rm{Mpc}^{-1}$. Fig.~\ref{fig:combinedone} displays the contours of $ S_{\delta^{(2)}}(k_1,k_2,k_3)$ with a scale-invariant GWs source on the left and a blue-tilted source on the right. The contours are plotted in the plane of $(k_1/k_3, k_2/k_3)$, where we have fixed the value of $k_3$ at $0.005 \text{ Mpc}^{-1}$, and arranged the momenta such that $k_1 \leq k_2 \leq k_3$. 
\subsection{Monochromatic source}
A monochromatic power spectrum of gravitational waves, characterized by a sharp peak at a specific wavenumber, serves as a useful approximation for models predicting localized enhancements in the tensor power spectrum. Such spectra are typically parameterized as 
\begin{equation}
\Delta_T^2(k) = A_T\, \delta \left(\ln \frac{k}{k_*}\right)\,,
\end{equation}
where \(A_T\) is the amplitude of the tensor spectrum, and \(k_*\) is the characteristic scale at which the peak occurs. Inflationary scenarios that predict such a spectrum are models involving parametric resonance \cite{Cai:2020ovp}. These models generate a highly localized enhancement in the gravitational wave spectrum, which could have significant observational implications for interferometers.
In Fig.~\ref{fig:Mono}, we show the bispectrum for the matter perturbation sourced by monochromatic GWs, at only $k_3 = k_*$ since we have the condition $k < 2k_*$ that comes from momentum conservation. 
\begin{figure}[t!]
    \centering
    \begin{subfigure}
             \centering
 \includegraphics[width=0.48\linewidth]{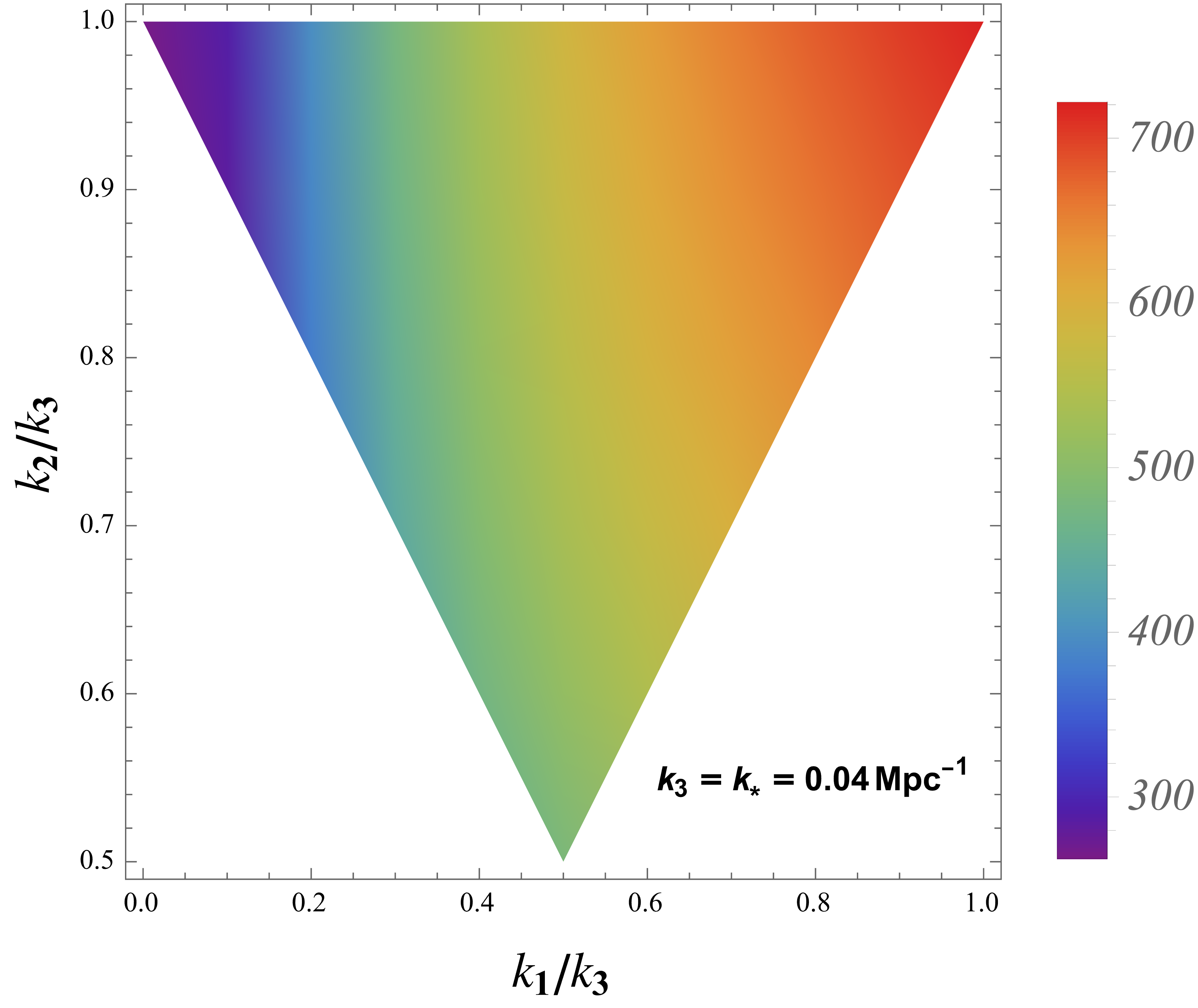}
    \end{subfigure}
    \hfill
    \begin{subfigure}
             \centering
\includegraphics[width=0.49\linewidth]{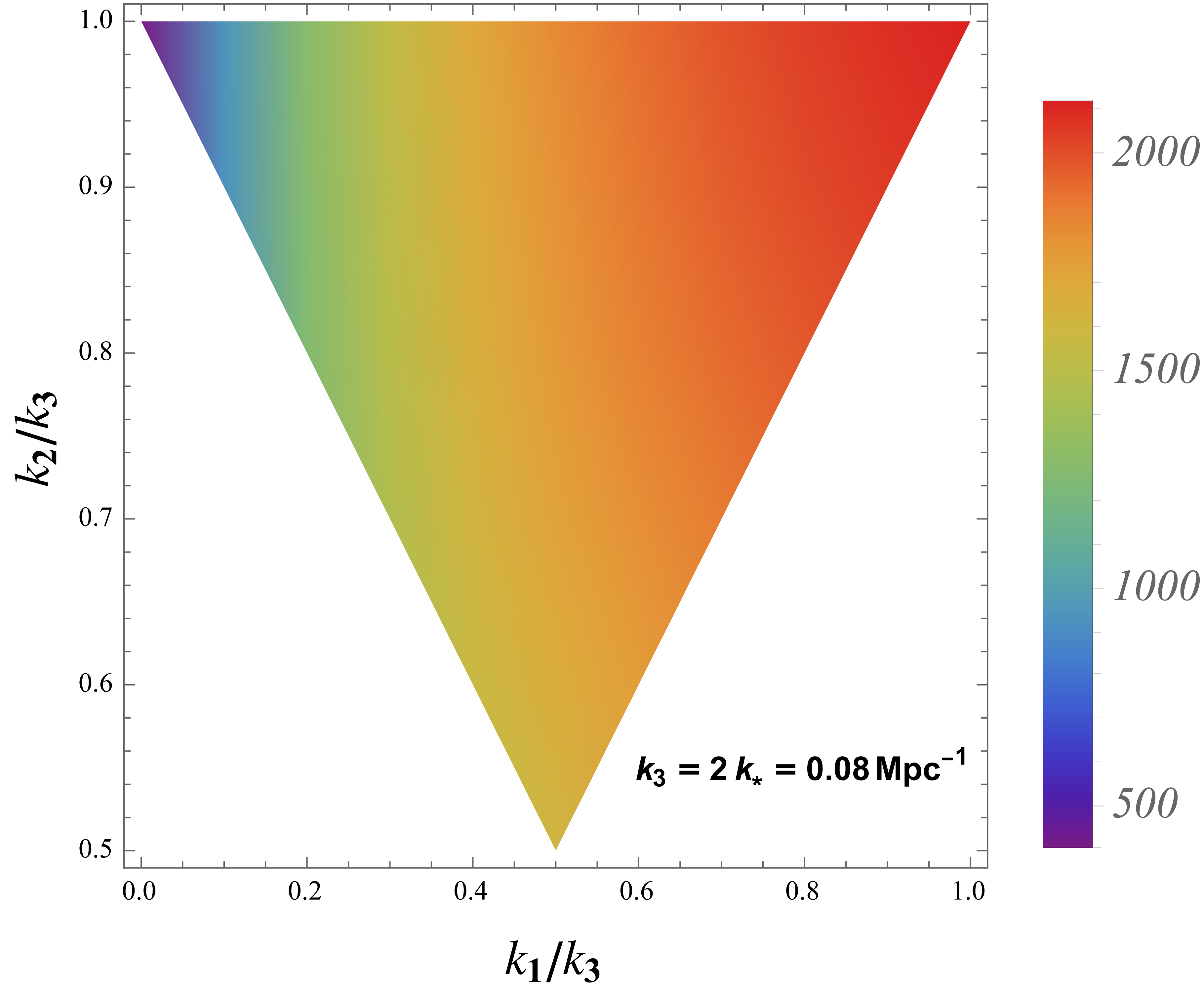}
    \end{subfigure}
    \caption{Normalized shapes for the bispectrum of the matter density contrast by a Gaussian bump GWs source with $k_3$ fixed at $k_3 = k_*$ (left) and $k_3 = 2k_*$ (right).}
    \label{fig:Gauss}
\end{figure}
\subsection{Gaussian-bump source}
Axion inflation models \cite{Namba:2015gja, Dimastrogiovanni:2016fuu}, where an axion-like field couples to SU(2) gauge fields as a spectator field during inflation, can generate a large Gaussian bump in the GWs spectrum. These models generate GWs with amplitudes comparable to or surpassing scalar perturbations, making them compelling targets for both CMB B-mode polarization observations and GWs interferometers \cite{Garcia-Bellido:2016dkw, Bartolo:2016ami,Maleknejad:2016qjz,Maleknejad:2018nxz}. The tensor power spectrum for the Gaussian bump is given by
\begin{equation}
\Delta_{T}^2(k) = A_{T} e^{-\frac{1}{2\sigma^2} \ln^2\left(\frac{k}{k_*}\right)}\,,
\end{equation}
where \( A_T \) is the amplitude, \( k_* \) is the scale where the bump is centred around, and \( \sigma \) is its width. Fig.~\ref{fig:Powerspectrum} shows the significant impact of the Gaussian bump source on the matter power spectrum. 
In Fig.~\ref{fig:Gauss}, we present the dimensionless shape of the bispectrum for the matter perturbation sourced by Gaussian bump GWs, with $k_3 = k_*$ on the left and $k_3 = 2k_*$ on the right.

\section{Discussion and conclusion}\label{sec6:Conc}
In this paper we have studied the intrinsic non-Gaussianity of tensor-induced scalar modes. Starting from the sourced equation for the dark matter density contrast, we derived the full bispectrum, confirming the significant non-Gaussianity of this effect, as previously highlighted in \cite{Bari:2021xvf}. Unlike their linear counterparts, these second-order matter perturbations vanish on super-horizon scales, leaving distinct signatures only on sub-horizon scales without affecting large-scale CMB anisotropies, though they may influence smaller scales. 

We have considered various primordial tensor power spectra as sources in our analysis, including scale-invariant, blue-tilted, Gaussian-bump, and monochromatic sources, while keeping the GWs parameters consistent with \cite{Bari:2021xvf}. We present the numerical results for the normalized bispectrum, keeping \( k_3 \) fixed and ordering the momenta as \( k_1 \leq k_2 \leq k_3 \). We plot $ S_{\delta^{(2)}}(k_1,k_2,k_3)$ for each of our selected GWs source models. We found that the bispectrum shape depends on the scale dependence of the input GWs model, suggesting that different inflationary models may be distinguished based on this bispectrum signature. In particular, we demonstrate that models with monochromatic GWs and more realistically, Gaussian-bump power spectra produce a strong bispectrum signal peaking in the equilateral configuration. This result aligns with expectations and mirrors the opposite effect, scalar-induced tensor modes, as shown in \cite{Espinosa:2018eve}. In both cases, sourcing occurs primarily around horizon crossing, and the bispectrum peaks when the sourcing momenta are equal in magnitude. This logic, not surprisingly, applies only to peaked sources, as we can observe that for power-law sources, our bispectrum peaks at squeezed and folded shapes.

The amplitude of the normalized bispectrum was \(\mathcal{O}(10^2)\) and \(\mathcal{O}(10^3)\) for the Gaussian-bump source for two cases where we fixed $k_3$ at the value of the peak and twice the peak scale of the tensor power spectrum, while for the monochromatic source, it reached \(\mathcal{O}(10^4)\), similarly to the scalar-induced case \cite{Espinosa:2018eve}, where a similar amplitude was found. 

This paper confirms that the tensor-induced matter perturbations would exhibit a high level of non-Gaussianity. The results presented in this paper open a complementary window for detecting and constraining PGWs on a range of scales that remain largely unconstrained. Future galaxy surveys including Euclid \cite{Euclid:2019clj}, DESI \cite{DESI:2018ymu}, SPHEREx \cite{SPHEREx:2016vbo}, SKA \cite{SKA:2018ckk}, the Roman Space Telescope \cite{Rose:2021nzt}, and the Vera Rubin Observatory (LSST) \cite{LSSTDarkEnergyScience:2021ryz} could provide a powerful means to detect or constrain this signature. Given the distinct origin of the non-Gaussianity studied here, estimating observational signatures for it would require a dedicated forecast tailored to this specific type of non-Gaussianity,  which we plan to pursue in future work.

\section*{Acknowledgments}
The authors acknowledge the useful discussions and comments received during the COSMO CLASSIC 2024 meeting in Asiago. P.B. is supported by IBS under the project code IBS-R018-D3. SM acknowledges partial financial support from the COSMOS
network (www.cosmosnet.it) through the ASI (Italian Space Agency) Grants 2016-24-H.0, 2016-24-H.1-2018 and 2020-9-HH.0.
\begin{appendices}
\numberwithin{equation}{section}

\section{Six-Point function contraction}\label{AppA}
Analogous to \cite{Espinosa:2018eve}, we consider the eight possible contractions of the six-point function of tensor perturbations in Eq. \eqref{threepnt}, with each contributing equally to the bispectrum thanks to the invariance of the bispectrum integral (equivalently Eq. \eqref{threepnt} here) under the exchange of the subscripts $1, 2$ and  $3$ and under $\bm{p}_i\rightarrow \bm{q}_i= \bm{k}_i-\bm{p}_i$.
Therefore, we can evaluate the three-point function for any of these configurations and multiply the result by $8$.
In our case, the three-point correlation function of GWs-induced matter perturbations also requires considering the exchange of polarization tensors.
\begin{align}\label{6pt_1}
\langle 16\rangle \langle 23\rangle \langle 45\rangle&= \langle A_{\sigma'_1} (\bm{p}_1)A_{\sigma_3} (\bm{q}_3)\rangle \langle A_{\sigma_1} (\bm{q}_1) A_{\sigma'_2} (\bm{p}_2)\rangle \langle A_{\sigma_2} (\bm{q}_2) A_{\sigma'_3} (\bm{p}_3)\rangle \,,  
\end{align}
\begin{align}
  \langle 13\rangle \langle 26\rangle \langle 45\rangle&=\langle A_{\sigma'_1} (\bm{p}_1)A_{\sigma'_2} (\bm{p}_2)\rangle  \langle A_{\sigma_1} (\bm{q}_1) A_{\sigma_3} (\bm{q}_3)\rangle \langle A_{\sigma_2} (\bm{q}_2) A_{\sigma'_3} (\bm{p}_3)\rangle \,,
\end{align}
obtained from Eq. \eqref{6pt_1} by $\bm{p}_1\rightarrow \bm{q}_1$, and $\sigma_1 \leftrightarrow \sigma'_1$.
\begin{align}
    \langle 15\rangle \langle 23\rangle \langle 46\rangle&=\langle A_{\sigma'_1} (\bm{p}_1)A_{\sigma'_3} (\bm{p}_3)\rangle \langle A_{\sigma_1} (\bm{q}_1) A_{\sigma'_2} (\bm{p}_2)\rangle \langle A_{\sigma_2} (\bm{q}_2) A_{\sigma_3} (\bm{q}_3)\rangle \,,
\end{align}
   obtained from Eq. \eqref{6pt_1} by $\bm{p}_3\rightarrow \bm{q}_3$, and $\sigma_3 \leftrightarrow \sigma'_3$. 
\begin{align}
    \langle 16\rangle \langle 24\rangle \langle 35\rangle&= \langle A_{\sigma'_1} (\bm{p}_1)A_{\sigma_3} (\bm{q}_3)\rangle \langle A_{\sigma_1} (\bm{q}_1)A_{\sigma_2} (\bm{q}_2) \rangle \langle A_{\sigma'_2} (\bm{p}_2) A_{\sigma'_3} (\bm{p}_3)\rangle \,, 
\end{align}
obtained from Eq. \eqref{6pt_1} by $\bm{p}_2\rightarrow \bm{q}_2$, and $\sigma_2 \leftrightarrow \sigma'_2$. 
 \begin{align}
\langle 14\rangle \langle 25\rangle \langle 36\rangle&= \langle A_{\sigma'_1} (\bm{p}_1)A_{\sigma_2} (\bm{q}_2)\rangle \langle A_{\sigma_1} (\bm{q}_1) A_{\sigma'_3} (\bm{p}_3)\rangle  \langle  A_{\sigma'_2} (\bm{p}_2)A_{\sigma_3} (\bm{q}_3)\rangle \,,  
\end{align}
obtained from Eq. \eqref{6pt_1} by changing subscripts $2 \leftrightarrow 3$. 
 \begin{align}
\langle 13\rangle \langle 25\rangle \langle 46\rangle&= \langle A_{\sigma'_1} (\bm{p}_1)A_{\sigma'_2} (\bm{p}_2)\rangle \langle A_{\sigma_1} (\bm{q}_1)A_{\sigma'_3} (\bm{p}_3)\rangle \langle A_{\sigma_2} (\bm{q}_2) A_{\sigma_3} (\bm{q}_3)\rangle    \,,  
\end{align}
obtained from Eq. \eqref{6pt_1} by $\bm{p}_1\rightarrow \bm{q}_1$,  $1 \leftrightarrow 2$, and $\sigma_2 \leftrightarrow \sigma'_2$.  
 \begin{align}
\langle 14\rangle \langle 26\rangle \langle 35\rangle&= \langle A_{\sigma'_1} (\bm{p}_1)A_{\sigma_2} (\bm{q}_2)\rangle  \langle A_{\sigma_1} (\bm{q}_1) A_{\sigma_3} (\bm{q}_3)\rangle \langle  A_{\sigma'_2} (\bm{p}_2)A_{\sigma'_3} (\bm{p}_3)\rangle \,,  
\end{align}
obtained from Eq. \eqref{6pt_1} by $\bm{p}_3\rightarrow \bm{q}_3$,  $1 \leftrightarrow 2$, and $\sigma_3 \leftrightarrow \sigma'_3$.  
\begin{align}
\langle 15\rangle \langle 24\rangle \langle 36\rangle&= \langle A_{\sigma'_1} (\bm{p}_1)A_{\sigma'_3} (\bm{p}_3)\rangle  \langle A_{\sigma_1} (\bm{q}_1) A_{\sigma_2} (\bm{q}_2) \rangle \langle A_{\sigma'_2} (\bm{p}_2)A_{\sigma_3} (\bm{q}_3)\rangle  \,,  
\end{align}
obtained from Eq. \eqref{6pt_1} by $\bm{p}_2\rightarrow \bm{q}_2$,  $1 \leftrightarrow 2$, and $\sigma_1 \leftrightarrow \sigma'_1$.

We compute Eq. \eqref{6pt_1} and get
\begin{align}\label{contr}
    \langle 16\rangle \langle 23\rangle \langle 45\rangle= (2\pi)^9\delta^3(\bm{p}_1+\bm{q}_3)\delta^3(\bm{q}_1+\bm{p}_2) \delta^3(\bm{q}_2+\bm{p}_3)\delta_{\sigma'_1 \sigma_3}\delta_{\sigma_1 \sigma'_2}\delta_{\sigma_2 \sigma'_3}
    P_{\mathcal{\sigma}_3}(p_1) P_{\mathcal{\sigma}_1}(p_2) P_{\mathcal{\sigma}_2}(p_3)\,,
\end{align}
where $P(p_1)(=(2\pi^2/p_1^3)\Delta^2(p_{1}))$ is the dimensional power spectrum of GWs.

\section{Contraction of polarization tensors}\label{AppB}
\subsection{Power Spectrum case}

We consider the coordinate system $\left\{\boldsymbol{\hat{e}}_{x}, \boldsymbol{\hat{e}}_{y}, \boldsymbol{\hat{e}}_{z}\right\}$ arbitrarily oriented and at rest with respect to the isotropic SGWB we analyze. In this reference frame, the $\mathbf{k}$ wave vector of an incoming plane GW sets the orthonormal basis.
\begin{equation}
    \boldsymbol{\hat{u}}(\boldsymbol{\hat{k}})=\frac{\boldsymbol{\hat{k}} \times \boldsymbol{\hat{e}}_{z}}{\left|\boldsymbol{\hat{k}} \times \boldsymbol{\hat{e}}_{z}\right|}\,, \quad \boldsymbol{\hat{v}}(\boldsymbol{\hat{k}})=\boldsymbol{\hat{k}} \times \boldsymbol{\hat{u}}\,,
\end{equation}
(where $\boldsymbol{\hat{k}}$ is unit vector in the direction of $\mathbf{k}$ and we will denote its magnitude by $k=|\mathbf{k}|)$ and we assume that $\boldsymbol{p}$ is a generic vector such that
\begin{align}
\boldsymbol{p}=(\sin \theta \cos \phi, \sin \theta \sin \phi, \cos \theta) \,.
\end{align}

We introduce to define the ``plus" $(+)$ and ``cross" $(\times)$ polarization tensors
\begin{equation}
    \epsilon_{a b}^{(+)}(\boldsymbol{\hat{k}})=\frac{\boldsymbol{\hat{u}}_{a} \boldsymbol{\hat{u}}_{b}-\boldsymbol{\hat{v}}_{a} \boldsymbol{\hat{v}}_{b}}{\sqrt{2}}\,, \quad \epsilon_{a b}^{(\times)}(\boldsymbol{\hat{k}})=\frac{\boldsymbol{\hat{u}}_{a} \boldsymbol{\hat{v}}_{b}+\boldsymbol{\hat{v}}_{a} \boldsymbol{\hat{u}}_{b}}{\sqrt{2}}  \,.
\end{equation}

The polarization tensors $\epsilon_{a b}^{(+)}$ and $\epsilon_{a b}^{(\times)}$ fulfill the conditions
\begin{equation}
    \begin{array}{lll}
\epsilon_{a b}^{+/ \times}(\boldsymbol{\hat{k}})=\epsilon_{a b}^{+/ \times *}(\boldsymbol{\hat{k}}), & \epsilon_{a b}^{+}(\boldsymbol{\hat{k}})=\epsilon_{a b}^{+}(-\boldsymbol{\hat{k}}), & \epsilon_{a b}^{\times}(-\boldsymbol{\hat{k}})=-\epsilon_{a b}^{\times}(\boldsymbol{\hat{k}}), \\
\epsilon_{a b}^{+}(\boldsymbol{\hat{k}}) \epsilon_{a b}^{+}(\boldsymbol{\hat{k}})=1, & \epsilon_{a b}^{\times}(\boldsymbol{\hat{k}}) \epsilon_{a b}^{\times}(\boldsymbol{\hat{k}})=1, & \epsilon_{a b}^{+}(\boldsymbol{\hat{k}}) \epsilon_{a b}^{\times}(\boldsymbol{\hat{k}})=0 .
\end{array}
\end{equation}

Now, we can write the expression in Eq. \eqref{f} as 
\begin{equation}
   f\big(k,p,\theta\big)\equiv  \sum_{\sigma_{1} \sigma_{1}^{\prime}} \epsilon_{i j}^{\sigma_{1}}\left(\boldsymbol{\hat{p}}\right) \epsilon_{i j}^{\sigma_{1}^{\prime}}\left(\widehat{\boldsymbol{k}-\boldsymbol{p}}\right) \epsilon_{k l}^{\sigma_{1} *}\left(\boldsymbol{\hat{p}}\right) \epsilon_{k l}^{\sigma_{1}{ }^{\prime} *}\left(\widehat{\boldsymbol{k}-\boldsymbol{p}}\right) \,.
\end{equation}

Using the previous properties, we know that the polarization tensor satisfies the following identity
\begin{align} \label{polid}
2 \sum_{\lambda} \epsilon_{i j, \lambda}(\boldsymbol{\hat{k}}) \epsilon_{a b, \lambda}^{*}(\boldsymbol{\hat{k}})= & \left(\delta_{i a}-\boldsymbol{\hat{k}}_{i} \boldsymbol{\hat{k}}_{a}\right)\left(\delta_{j b}-\boldsymbol{\hat{k}}_{j} \boldsymbol{\hat{k}}_{b}\right)+\left(\delta_{i b}-\boldsymbol{\hat{k}}_{i} \boldsymbol{\hat{k}}_{b}\right)\left(\delta_{j a}-\boldsymbol{\hat{k}}_{j} \boldsymbol{\hat{k}}_{a}\right) \\\nonumber&
-\left(\delta_{i j}-\boldsymbol{\hat{k}}_{i} \boldsymbol{\hat{k}}_{j}\right)\left(\delta_{a b}-\boldsymbol{\hat{k}}_{a} \boldsymbol{\hat{k}}_{b}\right) \,,
\end{align}

So the result of contraction reads 
\begin{equation}
    f\big(k,p,\theta\big)\equiv \frac{1}{4}\left[1+6\left(\boldsymbol{\hat{p}}\cdot\left(\boldsymbol{\widehat{k-p}}\right)\right)^{2}+\left(\boldsymbol{\hat{p}} \cdot\left(\boldsymbol{\widehat{k-p}}\right)\right)^{4}\right]\,.
\end{equation}

\begin{figure}[H]
    \centering
    \includegraphics[width=0.7\linewidth]{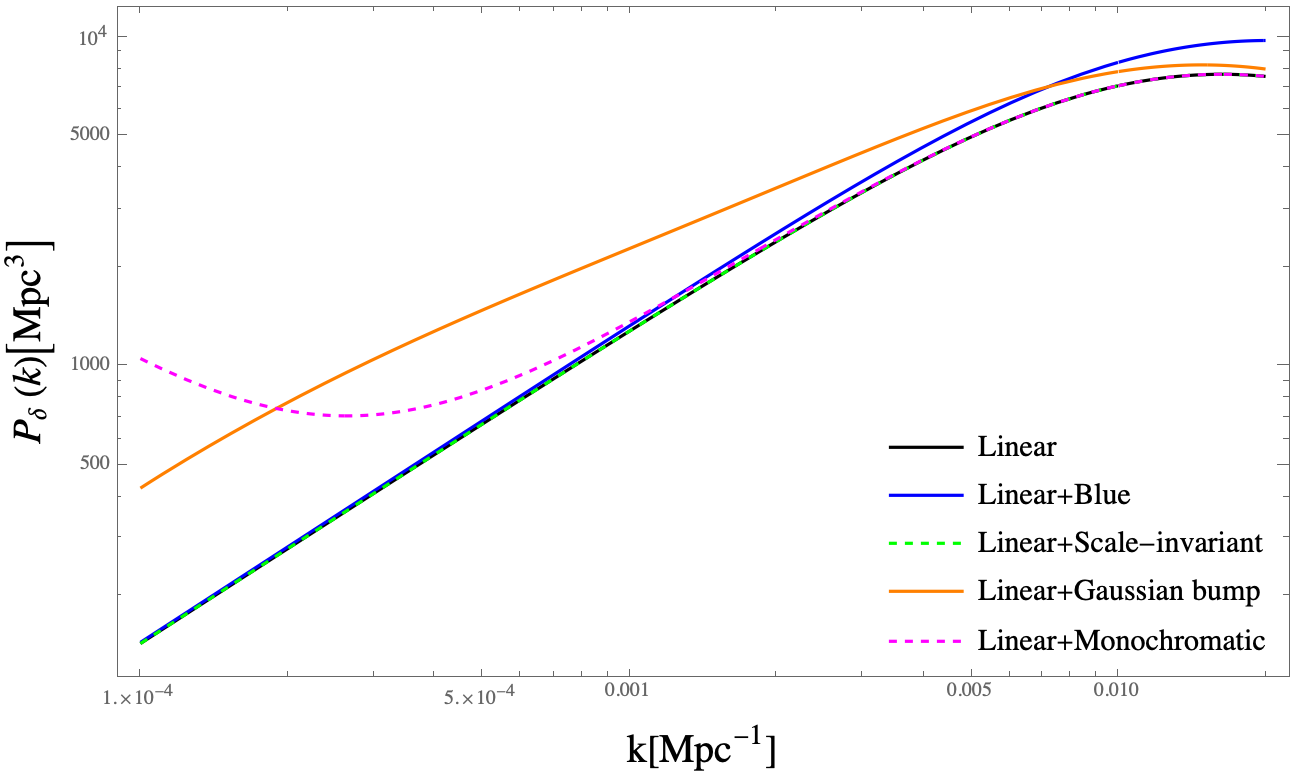}
    \caption{Impact of different GW sources on the matter power-spectrum; Blue-tilted: ($A_T = 1.26 \times 10^{-10}, \quad n_T = 0.32, \quad k_* = k_{\text{CMB}} = 0.01 \text{ Mpc}^{-1}$),  Monochromatic:  ($A_T = 10^{-5}, \quad k_* = 0.008 \text{ Mpc}^{-1}$),   Gaussian bump:  ($A_T = 10^{-5}, \quad \sigma = 2, \quad k_* = 0.04 \text{ Mpc}^{-1}$),   Scale-invariant: ($A_T = 1.26 \times 10^{-10} $). The plotted result represents a more accurate version of that in \cite{Bari:2021xvf}, achieved by increasing the numerical precision of the integration. }
\label{fig:Powerspectrum}
\end{figure}

\subsection{Bispectrum case}

The expression of polarization tensors in Eq. \eqref{bispec} is given by 
\begin{equation} \label{pol}
\sum_{\sigma_{1} \sigma_{2} \sigma_3} 
\epsilon_{ij}^{\sigma_{1}}\left(\boldsymbol{\hat{p}_{1}}\right) \epsilon^{\sigma_{2}^{*} ij}\left(\boldsymbol{\hat{p}_2}\right) \epsilon_{kl}^{\sigma_{2}}\left(\boldsymbol{\hat{p}_{2}}\right) \epsilon^{ \sigma_{3}^{*} kl}\left(\boldsymbol{\hat{p}_3}\right) \epsilon_{mn}^{\sigma_{3}}\left(\boldsymbol{\hat{p}_{3}}\right) \epsilon^{\sigma_{1}^{*} mn } \left(\boldsymbol{\hat{p}_1}\right)\,.
\end{equation}

Considering that the wave vector $\boldsymbol{k_3}$ is directed along the negative x-axis and defining the magnitudes of the momenta $\boldsymbol{k_i}$ as $k_1, k_2,$ and $k_3$, respectively. Furthermore, defining $\theta$ as the angle between $\boldsymbol{k_1}$ and the x-axis. Then, we obtain the following 
\begin{align}
  \cos{\theta} = \frac{k_1^2+k_3^2-k_2^2}{2 k_1 k_3}\,.
\end{align}

From Eq. \eqref{pi} we can get the unit vectors of the momenta $\boldsymbol{p}_i$ as following 
\begin{align} \label{piu}
\mathbf{\hat{p}}_1 &= \frac{1}{\sqrt{r^2 + \ell^2}} (r \cos \alpha, r \sin \alpha, \ell)\,,
\\\nonumber 
 \mathbf{\hat{p}}_2 &= \frac{1}{\sqrt{k_1^2 + r^2 
    - 2k_1 r \cos(\theta - \alpha) + \ell^2}} (-k_1 \cos{\theta}+r \cos \alpha, -k_1 \sin{\theta} + r \sin \alpha, \ell)\,,
\\\nonumber 
\mathbf{\hat{p}}_3 &= \frac{1}{\sqrt{r^2 - 2rk_3 \cos \alpha + k_3^2 + \ell^2}} (-k_3 + r \cos \alpha, r \sin \alpha, \ell)\,.
\end{align}

Again, with the identity Eq. \eqref{polid} defined in the previous section, we can solve the contraction in Eq. \eqref{pol} to obtain 
\begin{align} \label{contraction}
C &= \frac{1}{8} \left[-4 (\hat{\boldsymbol{p}}_1.\hat{\boldsymbol{p}}_2) (\hat{\boldsymbol{p}}_1.\hat{\boldsymbol{p}}_3) (\hat{\boldsymbol{p}}_2.\hat{\boldsymbol{p}}_3) +5(\hat{\boldsymbol{p}}_1.\hat{\boldsymbol{p}}_2)^2 (\hat{\boldsymbol{p}}_1.\hat{\boldsymbol{p}}_3)^2+(\hat{\boldsymbol{p}}_1.\hat{\boldsymbol{p}_2})^2 (\hat{\boldsymbol{p}}_1.\hat{\boldsymbol{p}}_3)^2 (\hat{\boldsymbol{p}}_2.\hat{\boldsymbol{p}}_3)^2 \right.\nonumber\\&\left.+5 (\hat{\boldsymbol{p}}_1.\hat{\boldsymbol{p}}_2)^2 (\hat{\boldsymbol{p}}_2.\hat{\boldsymbol{p}}_3)^2- 4 (\hat{\boldsymbol{p}}_1.\hat{\boldsymbol{p}}_2) (\hat{\boldsymbol{p}}_1.\hat{\boldsymbol{p}}_3) (\hat{\boldsymbol{p}}_2.\hat{\boldsymbol{p}}_3)^3 \right.\nonumber\\&\left. - 4 (\hat{\boldsymbol{p}}_1.\hat{\boldsymbol{p}}_2) (\hat{\boldsymbol{p}}_1.\hat{\boldsymbol{p}}_3)^3 (\hat{\boldsymbol{p}}_2.\hat{\boldsymbol{p}}_3) +4 (\hat{\boldsymbol{p}}_1.\hat{\boldsymbol{p}}_2) (\hat{\boldsymbol{p}}_1.\hat{\boldsymbol{p}}_3) (\hat{\boldsymbol{p}}_2.\hat{\boldsymbol{p}}_3)\right.\nonumber\\&\left.+5 (\hat{\boldsymbol{p}}_1.\hat{\boldsymbol{p}}_3)^2 (\hat{\boldsymbol{p}}_2.\hat{\boldsymbol{p}}_3)^2+2 (\hat{\boldsymbol{p}}_1.\hat{\boldsymbol{p}}_2)^4 + (\hat{\boldsymbol{p}}_1.\hat{\boldsymbol{p}}_2)^2+2 (\hat{\boldsymbol{p}}_1.\hat{\boldsymbol{p}}_3)^4+(\hat{\boldsymbol{p}}_1.\hat{\boldsymbol{p}}_3)^2 \right.\nonumber\\&\left. +2 (\hat{\boldsymbol{p}}_2.\hat{\boldsymbol{p}}_3)^4+(\hat{\boldsymbol{p}}_2.\hat{\boldsymbol{p}}_3)^2-1\right]\,,
\end{align}
which is used together with Eq. \eqref{piu} in Eq. \eqref{bispec} to solve the integral numerically. 
\end{appendices}
\newpage
\bibliographystyle{ieeetr}
\bibliography{PGWinduceddelta.bib}
\end{document}